# A path-specific effect approach to mediation analysis with time-varying mediators and time-to-event outcomes accounting for competing risks

Arce Domingo-Relloso,[1,2] Yuchen Zhang,[1] Ziqing Wang,[1] Astrid M Suchy-Dicey,[3] Dedra S Buchwald,[4] Ana Navas-Acien,[2] Joel Schwartz,[5] Kiros Berhane,[1] Brent A. Coull,[6] Linda Valeri[1,7]

[1] Department of Biostatistics, Columbia University Mailman School of Public Health, New York, NY

[2] Department of Environmental Health Sciences, Columbia University Mailman School of Public Health, New York, NY

[3] Huntington Medical Research Institutes, Pasadena, CA

[4] Neurosciences Institute, Department of Neurological Surgery, University of Washington, Seattle, WA

[5] Department of Environmental Health Sciences, Harvard TH Chan School of Public Health, Cambridge, MA

[6] Department of Biostatistics, Harvard TH Chan School of Public Health, Boston, MA

[7] Department of Epidemiology, Harvard TH Chan School of Public Health, Boston, MA

**Abstract**

Not accounting for competing events in survival analysis can lead to biased estimates, as individuals who die from other causes do not have the opportunity to develop the event of interest. Formal definitions and considerations for causal effects in the presence of competing risks have been published, but not for the mediation analysis setting. We propose, for the first time, an approach based on the path-specific effects framework to account for competing risks in longitudinal mediation analysis with time-to-event outcomes. We do so by considering the pathway through the competing event as another mediator, which is nested within our longitudinal mediator of interest. We provide a theoretical formulation and related definitions of the effects of interest based on the mediational g-formula, as well as a detailed description of the algorithm. We also present an application of our algorithm to data from the Strong Heart Study, a prospective cohort of American Indian adults. In this application, we evaluated the mediating role of the blood pressure trajectory (measured during three visits) on the association between arsenic and cadmium, in separate models, with time to cardiovascular disease, accounting for competing risks by death. Identifying the effects through different paths enables us to evaluate the impact of metals on the outcome of interest, as well as through competing risks, more transparently.

## 1. Introduction

In survival settings, competing events refer to any event that makes it impossible for the event of interest to occur. For example, death is a competing event for any other event, as when an individual dies, he/she cannot experience any further events. Not accounting for competing events in survival analysis can lead to biases caused by the fact that individuals that die from other causes do not have the opportunity to develop the event of interest (1). This has been clearly shown, for example, in the context of aging-related diseases, such as dementia. As reported in Rojas-Saunero et al. (2), counterintuitive findings have been observed when factors such as smoking (3) or cancer history (3) showed protective effects against dementia when ignoring death. This does not mean that those factors are protective of the dementia disease but because smokers and cancer patients tend to die earlier, this premature mortality makes them less likely to develop dementia. There is urgent need to develop new frameworks to effectively acknowledge competing risks, and to report causal effects that would take their potential impact into account.

Various statistical estimands have been proposed for competing risks in failure-time settings including marginal cumulative incidence, cause-specific cumulative incidence, marginal hazard, sub-distribution hazard and cause-specific hazard (4). However, Young et al. (5) was the first paper to present a formal definition of causal effects in the presence of competing risks using counterfactuals. In this paper, the authors describe two different estimands that can be considered in the presence of competing risks: the direct effect, defined as the risk under elimination of competing events, and the total effect, defined as the risk without elimination of competing events. Martinussen and Stensrud (6) propose to address the issue of competing events by considering a hypothetical scenario in which we have two separable treatments, the first only affects the event of interest and the second only affects the competing event.

In this work, we propose a framework for the causal mediation setting in presence of mediators measured repeatedly over time. We do this by considering the competing event as an additional mediator that is affected by past values of the mediator of interest (i.e. is nested with our mediator of interest), and in turn affects future values of both the mediator of interest and the outcome deterministically. This path specific effect structure has previously been described in the literature in the context of nested mediators (i.e. not in the context of competing risks) (7), but the authors did not consider time-varying nested mediators in that work. One of the key differences between our approach and previously considered approaches to deal with competing events is that we do not consider the framework in which causal effects are calculated under elimination of the competing event nor a framework in which the exposure can be separated into components that only affect the terminal event, as these scenarios are not realistic in many settings in which competing events by death simply cannot be eliminated or separated.

The paper is organized as follows: in section 2, we introduce the data structure and notation related to our causal setting. In section 3, we introduce the counterfactual estimands of causal mediation analysis in the presence of competing events. In section 4, we describe the identifiability assumptions. Section 5 describes the estimation of the effects. Section 6 is a full description of our algorithm. Section 7 includes an application of the approach to the analysis of the Strong Heart Study, a prospective cohort study of American Indian adults. This analysis evaluated the potential mediated effect of blood pressure on the association between metals and cardiovascular disease incidence, accounting for competing events by death. Last, section 8 includes discussion.

## 2. Data structure and notation

Consider $i = 1, \dots, n$ individuals exposed to different levels of a continuous exposure $A$ at baseline. Individuals are assumed to be independent and identically distributed. Let $M_{i1}, \dots, M_{iK}$ denote longitudinal measurements of a mediator for the $i$-th individual at visits $1, \dots, K$. For simplicity, we denote this vector as $M_1, \dots, M_K$ hereinafter. In addition, let $L_0$ represent a vector of baseline covariates, and $Y_k$, $D_k$ and $S_k$ denote indicators of the event of interest (e.g., cardiovascular disease), a competing event (e.g., death from all causes) and survival by time interval $k$, respectively. Our causal framework is depicted in the directed acyclic graph shown in Figure 1, provided for $K = 3$ even though it could easily be extended to allow for more visits. By definition, $D_0 \equiv Y_0 \equiv 0$ and $S_0 \equiv 1$, because the study population is restricted to those who have not yet experienced the event of interest or the competing event at baseline. We assume, without loss of generality, the temporal ordering $(A, M_k, D_k, S_k, Y_k)$ within each follow-up visit $k > 0$. For simplicity, we also assume that all variables are measured without error.

**Figure 1.** Directed Acyclic Graph for $K = 3$.

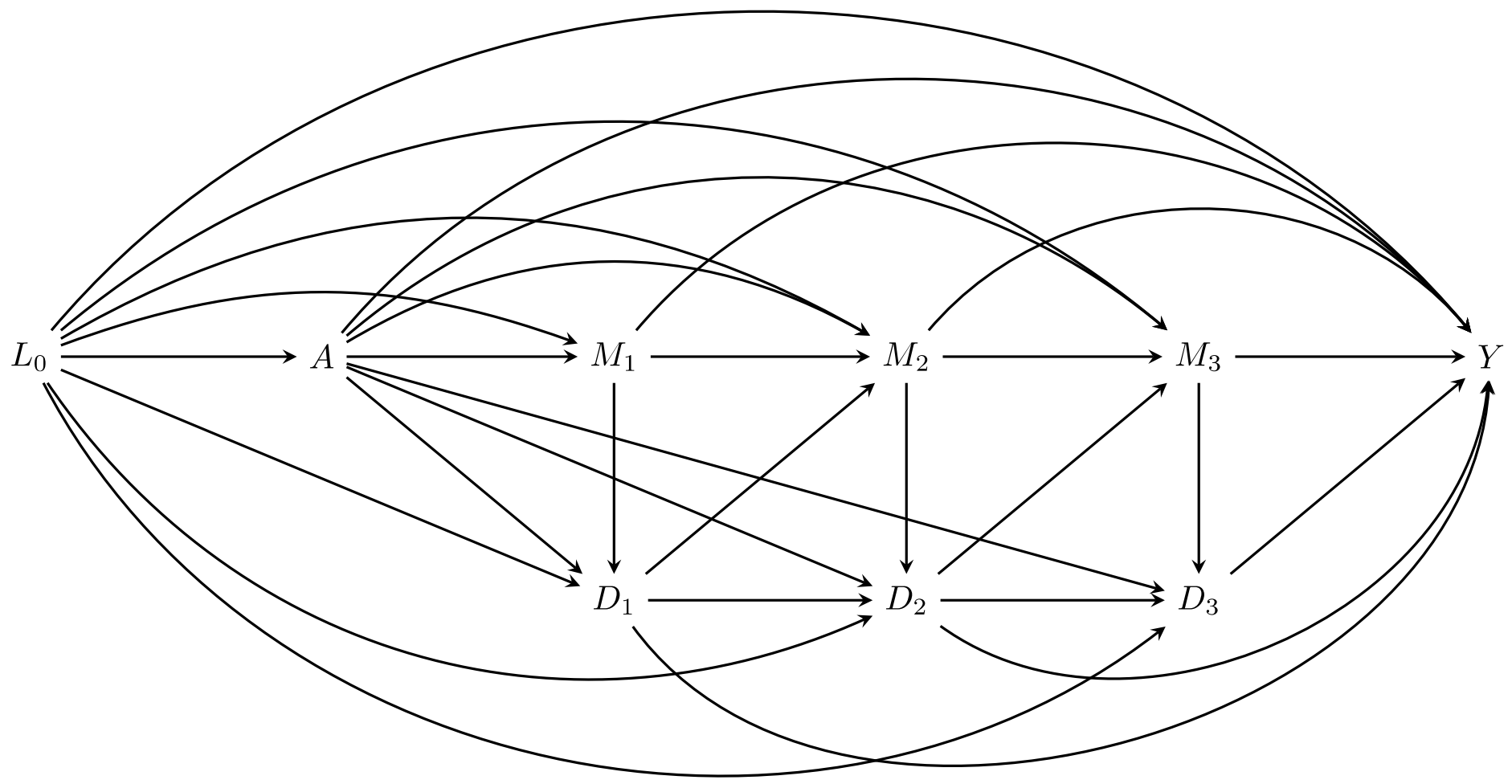

$A$ represents the exposure, $L_0$ represents a vector of baseline confounders, $M$ is the mediator measured at three time points $(M_1, M_2, M_3)$, $D$ is the competing risk indicator measured at three time points $(D_1, D_2, D_3)$ and $Y$ is the time-to-event outcome. Please note there is no arrow from $M_1$ and $M_2$ to $D_3$, from $M_1$ to $M_3$, from $M_1$ to $D_2$ and from $D_1$ to $D_3$ because we assume that our process is Markovian and therefore each longitudinal value of $M$ and $D$ is only affected by immediately prior values of $M$ and $D$. This assumption can be relaxed within the proposed approach.

Following the notation used in Young et al. (5), we denote the history of a random variable using overbars, e.g., $\overline{M}_k = (M_1, \ldots, M_k)$ is the history of the mediator of interest through interval $k$. Note that, if an individual is known to experience the competing event by visit $k > 0$ without having experienced the event of interest $(Y_{k-1} = 0, D_k = 1)$, then all future indicators of the event of interest will be observed and will be deterministically zero because, by definition, individuals that experience the competing event cannot experience the event of interest. Following Figure 1, the competing event indicator at a certain time point $k$ $(D_k)$ can be considered as another mediator which is nested within the vector of mediators $M$, in the sense that it is influenced by past values of the mediator of interest and in turn influences future values of the mediator of interest (in a deterministic way, because if an individual dies, the mediator is death-truncated).

## 3. Counterfactual estimands of causal mediation analysis in presence of competing events

We used the Lin-Ying additive hazards model (8), a special case of the Aalen additive hazards model with time-invariant coefficients, to model our time-to-event outcome. This model allows us to define our counterfactual outcome in terms of absolute attributable risks, but also in terms of survival probability. The additive hazards model has been described in the context of mediation analysis in previous work (9). For times $k = 0, \dots, K$, consider the counterfactual history of the mediator of interest $\bar{M}_{k+1}(a, \bar{S}_k(a))$, which refers to the value the history of the mediator would take had the individual, possibly contrary to fact, been exposed to $A = a$ and to the survival status $\bar{S}_k(a)$. Consider also the counterfactual history of the survival status $\bar{S}_k(a, \bar{M}_k(a))$, and the counterfactual outcomes $\gamma\left(a, \bar{M}_{k+1}(a, \bar{S}_k(a)), \bar{S}_k(a, \bar{M}_k(a))\right)$, which refers to the number of cases attributable to the exposure per a number of person-years (typically set to 10,000 or 100,000) at time $k$; and $P\left(Y_{k+1}\left(a, \bar{M}_k(a, \bar{S}_{k-1}(a)), \bar{S}_k(a, \bar{M}_k(a))\right) = 0\right)$, which refers to the probability of being free of the outcome at time $k + 1$.

We propose to decompose our causal mediation effects of interest in a path-specific effects approach to account for competing risks. Therefore, we define, for certain levels $a$ and $a^*$ of the exposure of interest, the following path-specific effects on the rate difference, as well as survival probability difference, scales:

1) Direct effect of the exposure on the outcome (DE): effect that operates neither through mortality nor through the longitudinal mediator of interest $M$ (see Supplementary Figure 1).

$$\gamma\left(a^{*},\bar{M}_k\big(a,\bar{S}_{k-1}(a)\big),\bar{S}_k\big(a,\bar{M}_k(a)\big)\right)-\gamma\left(a,\bar{M}_k\big(a,\bar{S}_{k-1}(a)\big),\bar{S}_k\big(a,\bar{M}_k(a)\big)\right),$$

being $\gamma$ the rate obtained from an additive hazards model, and

$$P\left(Y_{k+1}\left(a^{*},\bar{M}_k\big(a,\bar{S}_{k-1}(a)\big),\bar{S}_k\big(a,\bar{M}_k(a)\big)\right)=0\right)$$
$$-P\left(Y_{k+1}\left(a,\bar{M}_k\big(a,\bar{S}_{k-1}(a)\big),\bar{S}_k\big(a,\bar{M}_k(a)\big)\right)=0\right),$$

2) Indirect effect through the history of the mediator of interest (IEM), when the death process behaves under the reference level of exposure (Supplementary Figure 2).

$$\gamma\left(a^{*},\bar{M}_k\big(a^{*},\bar{S}_{k-1}(a)\big),\bar{S}_k\big(a,\bar{M}_k(a^{*})\big)\right)-\gamma\left(a^{*},\bar{M}_k\big(a,\bar{S}_{k-1}(a)\big),\bar{S}_k\big(a,\bar{M}_k(a)\big)\right),\text{ or}$$

$$P\left(Y_{k+1}\left(a^{*},\bar{M}_k\big(a^{*},\bar{S}_{k-1}(a)\big),\bar{S}_k\big(a,\bar{M}_k(a^{*})\big)\right)=0\right)$$
$$-P\left(Y_{k+1}\left(a^{*},\bar{M}_k\left(a,\bar{S}_{k-1}(a)\right),\bar{S}_k\left(a,\bar{M}_k(a)\right)\right)=0\right)$$

This is the effect that operates by impacting the mediator of interest directly, which in turn might impact the outcome (i) directly or (ii) via the mortality process. Please note that separating (i) and (ii) is impossible unless we make additional modeling (parametric) assumptions (see Discussion section).

3) Indirect effect through the history of the competing event (IED): effect that operates by impacting the death process directly, which has a deterministic relationship with the mediator of interest and the outcome, in the sense that the mediator is unobserved, and the outcome does not happen, if death happens before (Supplementary Figure 3).

$$\gamma\left(a^{*},\bar{M}_k\big(a^{*},\bar{S}_{k-1}(a^{*})\big),\bar{S}_k\big(a^{*},\bar{M}_k(a^{*})\big)\right)-\gamma\left(a^{*},\bar{M}_k\big(a^{*},\bar{S}_{k-1}(a)\big),\bar{S}_k\big(a,\bar{M}_k(a^{*})\big)\right),\text{ or}$$

$$P\left(Y_{k+1}\left(a^{*},\bar{M}_k\big(a^{*},\bar{S}_{k-1}(a^{*})\big),\bar{S}_k\big(a^{*},\bar{M}_k(a^{*})\big)\right)=0\right)$$
$$-P\left(Y_{k+1}\left(a^{*},\bar{M}_k\big(a^{*},\bar{S}_{k-1}(a)\big),\bar{S}_k\big(a,\bar{M}_k(a^{*})\big)\right)=0\right)$$

4) Total effect (TE):

$$\gamma\left(a^*, \bar{M}_k(a^*, \bar{S}_{k-1}(a^*)), \bar{S}_k(a^*, \bar{M}_k(a^*))\right) - \gamma\left(a, \bar{M}_k(a, \bar{S}_{k-1}(a)), \bar{S}_k(a, \bar{M}_k(a))\right), \text{ or}$$

$$P\left(Y_{k+1}\left(a^*, \bar{M}_k(a^*, \bar{S}_{k-1}(a^*)), \bar{S}_k(a^*, \bar{M}_k(a^*))\right) = 0\right)$$
$$- P\left(Y_{k+1}\left(a, \bar{M}_k\left(a, \bar{S}_{k-1}(a)\right), \bar{S}_k\left(a, \bar{M}_k(a)\right)\right) = 0\right)$$

Sum of direct effect and indirect effect through the mediator of interest (Supplementary Figure 4):

$$\gamma\left(a^*, \bar{M}_k(a^*, \bar{S}_{k-1}(a)), \bar{S}_k(a, \bar{M}_k(a^*))\right) - \gamma\left(a, \bar{M}_k\left(a, \bar{S}_{k-1}(a)\right), \bar{S}_k\left(a, \bar{M}_k(a)\right)\right), \text{ or}$$

$$P\left(Y_{k+1}\left(a^*, \bar{M}_k(a^*, \bar{S}_{k-1}(a)), \bar{S}_k(a, \bar{M}_k(a^*))\right) = 0\right)$$
$$- P\left(Y_{k+1}\left(a, \bar{M}_k\left(a, \bar{S}_{k-1}(a)\right), \bar{S}_k\left(a, \bar{M}_k(a)\right)\right) = 0\right)$$

Please note that the sum of the three path-specific effects leads to the total effect:

$$TE = DE + IEM + IED$$

For the rate difference, the effects can be identified using the estimated coefficients of the additive hazards model. The survival probability is calculated using the predict() function of the *timereg* R package, fixing a certain time point.

Also note that our definition of the total effect, $\mathbb{E}\left(Y_{k+1}(a^*, \bar{S}_k(a^*))\right) - \mathbb{E}\left(Y_{k+1}(a, \bar{S}_k(a))\right)$, is different than the definition of the total effect in traditional analysis that considers death as a censoring event, $\mathbb{E}(Y_{k+1}(a^*)|\bar{S}_k = 1) - \mathbb{E}(Y_{k+1}(a)|\bar{S}_k = 1)$, in the sense that the traditional total effect calculates the effect conditioning on survival up to time $k$, whereas the total effect accounting for competing risks calculates the effect when setting survival to the level it would take at the level of exposure $a$ versus $a^*$. Conversely, the sum of the direct effect and the indirect effect through M, $\mathbb{E}(Y_{k+1}(a^*,\bar{S}_k(a)) - \mathbb{E}(Y_{k+1}(a,\bar{S}_k(a))$, does not condition on survival, but sets survival to the level it would take under the reference level of exposure. The differences between these effects are also illustrated in our analysis of the Strong Heart Study data in section 8.

## 4. Identifiability assumptions

The following assumptions need to be fulfilled for the effects described in the previous section to be identifiable.

### 4.1. *Exchangeability*

4.1.1. No unmeasured confounding of the relationship between the exposure and the history of the outcome: $\bar{Y}_{k+1}(a,\bar{m},\bar{s}=1) \perp A|L_0$.

4.1.2. No unmeasured confounding of the relationship between the history of the mediator and both the outcome and the survival status at time $k+1$: $\big(Y_{k+1}(a^*,\bar{s}_k = 1,\bar{m}_k),\bar{S}_k(a^*,\bar{m}_k)\big) \perp \bar{M}_k|A,C,L_0$.

4.1.3. No unmeasured confounding of the relationship between the mediator of interest and the survival status at each time point: $\bar{S}_{k+1}(a,\bar{m}) \perp \bar{M}_{k+1}(a,\bar{s}=1)|A=a,\bar{M}_k = \bar{m}_k,\bar{Y}_k = 0,\bar{S}_k = 1,L_0$.

4.1.4. No unmeasured confounding of the relationship between the history of the mediator and the exposure: $\bar{M}_{k+1}(a,\bar{s}=1) \perp A|\, L_0,\bar{Y}_k = 0,\bar{S}_k = 1$.

4.1.5. No unmeasured confounding of the relationship between the history of the competing event and the exposure: $\bar{S}_{k+1}(a,\bar{m}) \perp A|\, L_0,\bar{Y}_k = 0,\bar{S}_k = 1$.

4.1.6. Cross-world assumption: no confounders of the mediators-outcome relationship affected by the exposure. $Y_{k+1}(a, \bar{m}, \bar{s} = 1) \perp \left(\bar{M}_k(a^*, \bar{s} = 1), \bar{S}_k(a^*, \bar{m})\right) | \bar{Y}_k = 0, \bar{S}_{k-1} = 1, A = a, L_0$.

Please note that the cross-world assumption would be violated if time-varying confounders exist. However, our approach can be extended to the setting in which we can consider time-varying confounders by adding an additional path to the path specific effects involving the time-varying confounder (see the discussion section).

*4.2.* <u>*Positivity:*</u> $P(A = a | L_0 = l_0) > 0$, and

$P(\bar{M}_{k+1} = \bar{m}_{k+1}, \bar{S}_{k+1} = \bar{s}_{k+1} | A = a, L_0 = l_0, \bar{Y}_k = 0, \bar{S}_k = 1) > 0, \forall k = 1, \dots, K.$

*4.3.* <u>*Consistency:*</u> If $A = a$ and $S_k = 1$, then, $\bar{M}_{k+1} = \bar{M}_{k+1}(a, \bar{s} = 1)$, $S_{k+1} = \bar{S}_{k+1}(a, \bar{m})$, $\bar{Y}_{k+1} = \bar{Y}_{k+1}(a, \bar{m}, \bar{s} = 1)$.

We additionally assume the stable unit treatment value assumption, i.e., that there is no interference across units.

Under these identifiability conditions, the expected value of the outcome at time $k + 1$, $\mathbb{E}[Y_{k+1}(a, \bar{M}_k, \bar{S}_k)]$ is identified by the following function of the observed data:

$$\phi(a, a, a) = \int_{\bar{m}_k} \mathbb{E}(Y_{k+1} | A = a, \bar{S}_k = 1, \bar{M}_k = \bar{m}_k, C = c) f(\bar{S}_k = 1 | A = a, \bar{M}_k = \bar{m}_k, C = c) f(\bar{M}_k = \bar{m}_k | A = a, \bar{S}_{k-1} = 1, C = c) d\bar{m}_k$$

This expression is called the mediational g-formula (10–12).

## 5. Effect estimation

Our path-specific effects, as described in section 3, can be computed using the mediational g-formula:

1) Direct effect of the exposure on the outcome:

$$\phi(a^*, a, a) - \phi(a, a, a)$$
$$= \int_{\bar{m}_k} \mathbb{E}(Y_{k+1} | A = a^*, \bar{S}_k = 1, \bar{M}_k = \bar{m}_k, C = c)\, f(\bar{S}_k$$
$$= 1 |\, A = a, \bar{M}_k = \bar{m}_k, C = c)\, f(\bar{M}_k = \bar{m}_k | A = a, \bar{S}_{k-1} = 1, C = c) d\bar{m}_k$$
$$- \int_{\bar{m}_k} \mathbb{E}(Y_{k+1} | A = a, \bar{S}_k = 1, \bar{M}_k = \bar{m}_k, C = c) f(\bar{S}_k$$
$$= 1 | A = a, \bar{M}_k = \bar{m}_k, C = c) f(\bar{M}_k = \bar{m}_k | A = a, \bar{S}_{k-1} = 1, C = c) d\bar{m}_k$$

2) Indirect effect through the history of the mediator of interest:

$$\phi(a^*, a^*, a^*) - \phi(a^*, a^*, a)$$
$$= \int_{\bar{m}_k} \mathbb{E}(Y_{k+1} | A = a^*, \bar{S}_k = 1, \bar{M}_k = \bar{m}_k, C = c) f(\bar{S}_k$$
$$= 1 | A = a^*, \bar{M}_k = \bar{m}_k, C = c) f(\bar{M}_k = \bar{m}_k | A = a^*, \bar{S}_{k-1} = 1, C = c) d\bar{m}_k$$
$$- \int_{\bar{m}_k} \mathbb{E}(Y_{k+1} | A = a^*, \bar{S}_k = 1, \bar{M}_k = \bar{m}_k, C = c) f(\bar{S}_k$$
$$= 1 | A = a^*, \bar{M}_k = \bar{m}_k, C = c) f(\bar{M}_k = \bar{m}_k | A = a, \bar{S}_{k-1} = 1, C = c) d\bar{m}_k$$

3) Indirect effect through the history of the competing event:

$$
\begin{aligned}
&\phi(a^*, a^*, a) - \phi(a^*, a, a) \\
&\quad = \int_{\bar{m}_k} \mathbb{E}(Y_{k+1} | A = a^*, \bar{S}_k = 1, \bar{M}_k = \bar{m}_k, C = c) f(\bar{S}_k \\
&\quad = 1 | A = a^*, \bar{M}_k = \bar{m}_k, C = c) f(\bar{M}_k = \bar{m}_k | A = a, \bar{S}_{k-1} = 1, C = c) d\bar{m}_k \\
&\quad - \int_{\bar{m}_k} \mathbb{E}(Y_{k+1} | A = a^*, \bar{S}_k = 1, \bar{M}_k = \bar{m}_k, C = c) f(\bar{S}_k \\
&\quad = 1 | A = a, \bar{M}_k = \bar{m}_k, C = c) f(\bar{M}_k = \bar{m}_k | A = a, \bar{S}_{k-1} = 1, C = c) d\bar{m}_k
\end{aligned}
$$

4) Total effect:

$$
\begin{aligned}
&\phi(a^*, a^*, a^*) - \phi(a, a, a) \\
&\quad = \int_{\bar{m}_k} \mathbb{E}(Y_{k+1} | A = a^*, \bar{S}_k = 1, \bar{M}_k = \bar{m}_k, C = c) f(\bar{S}_k \\
&\quad = 1 | A = a^*, \bar{M}_k = \bar{m}_k, C = c) f(\bar{M}_k = \bar{m}_k | A = a^*, \bar{S}_{k-1} = 1, C = c) d\bar{m}_k \\
&\quad - \int_{\bar{m}_k} \mathbb{E}(Y_{k+1} | A = a, \bar{S}_k = 1, \bar{M}_k = \bar{m}_k, C = c) f(\bar{S}_k \\
&\quad = 1 | A = a, \bar{M}_k = \bar{m}_k, C = c) f(\bar{M}_k = \bar{m}_k | A = a, \bar{S}_{k-1} = 1, C = c) d\bar{m}_k
\end{aligned}
$$

The proofs of these expressions are presented in the Supplementary Material.

## 6. Mediational g-formula algorithm with time-varying mediators and competing risks by death

Based on the DAG presented in Figure 1, our mediational g-formula algorithm proceeds as follows:

1. Choose two exposure values $a$ and $a^*$ (for example, percentiles 25th and 75th)
2. Fit sequential parametric linear and logistic regression models, respectively, for each time point $k = 1, \ldots, K$ for both $M$ and $D$: $M_k \sim M_{k-1} + A + L_0$, $D_k \sim M_k + A + L_0$.

3. Fit a parametric time-to-event additive hazards model for the outcome $Y$ using a counting process format (13) to adjust for $M$ in all time points: $Y \sim M + A + L_0$. We use the tmerge function from the *survival* R package to transform the database to a counting process format, and the *timereg* R package to fit a Lin-Ying additive hazards model.
4. Predict the counterfactual values of $M_k$ and $D_k$ for each iteration and each time point $k$: $M_k(a) = M_k\big(a, M_{k-1}(a), D_{k-1}(a)\big)$; $M_k(a^*) = M_k\big(a^*, M_{k-1}(a^*), D_{k-1}(a^*)\big)$; $M_k(a, a^*) = M_k\big(a, M_{k-1}(a), D_{k-1}(a^*)\big)$, and $D_k(a) = D_k\big(a, M_k(a), D_{k-1}(a)\big)$; $D_k(a^*) = M_k\big(a^*, M_k(a^*), D_{k-1}(a^*)\big)$; $D_k(a^*, a) = D_k\big(a^*, M_{k-1}(a), D_{k-1}(a^*)\big)$.
The death status of each participant is predicted by generating random values from a binomial distribution with the predicted probabilities of the logistic regression model. Thus, different individuals will be predicted to die/survive for different counterfactual scenarios.
5. Predict the counterfactual values of the outcome $Y$:
   - $Y(a) = Y\left(a, \bar{M}\big(a, \bar{D}(a)\big), \bar{D}\big(a, \bar{M}(a)\big)\right)$
   - $Y(a^*) = Y\left(a^*, \bar{M}\big(a^*, \bar{D}(a^*)\big), \bar{D}\big(a^*, \bar{M}(a^*)\big)\right)$
   - $Y\big(a^*, \bar{M}(a), \bar{D}(a)\big) = Y\left(a^*, \bar{M}\big(a, \bar{D}(a)\big), \bar{D}\big(a, \bar{M}(a)\big)\right)$
   - $Y\big(a^*, \bar{M}(a, a^*), \bar{D}(a^*, a)\big) = Y\left(a^*, \bar{M}\big(a, \bar{D}(a^*)\big), \bar{D}\big(a^*, \bar{M}(a)\big)\right)$.
6. Calculate the effects of interest for each iteration:
   - Direct effect: $\mathbb{E}\big[Y\big(a^*, M(a), D(a)\big)\big] - \mathbb{E}[Y(a)]$.
   - Indirect effect through M: $\mathbb{E}[Y(a^*)] - \mathbb{E}\big[Y\big(a^*, M(a, a^*), D(a^*, a)\big)\big]$.
   - Indirect effect through D: $\mathbb{E}\big[Y\big(a^*, M(a, a^*), D(a^*, a)\big)\big] - \mathbb{E}\big[Y\big(a^*, M(a), D(a)\big)\big]$.

- Total effect: $\mathbb{E}[Y(a^*)] - \mathbb{E}[Y(a)]$.

7. Use a quantile bootstrap procedure to calculate confidence intervals. Take the 50$^{th}$ percentile as the effect of interest, and the 2.5$^{th}$ and 97.5$^{th}$ percentiles as the 95 % lower and upper confidence intervals.

Please note that we make a Markovian assumption and only adjust each model for the mediator in the preceding time point to avoid multicollinearity. These assumptions can be relaxed, if needed. Our models allow for exposure-mediator interactions and for non-linear effects.

## 7. Application: metals, blood pressure and cardiovascular disease in the Strong Heart Study

We evaluated the potential mediating role of the systolic blood pressure trajectory on the association between arsenic and cadmium (in separate models) and incident CVD, accounting for competing risks by death. Both arsenic and cadmium have shown to be associated with elevated blood pressure (14,15) and with incident CVD (16,17) in the SHS.

*Study population: the Strong Heart Study*

A total of 4,549 men and women aged 45-74 years were recruited for the SHS from 13 American Indian tribes across three study centers in South Dakota, North Dakota, Oklahoma and Arizona with baseline visits occurring between 1989 and 1991 (18). In 2016, one tribal community (N=1,033) withdrew consent to use their data for research, leaving 3,516 participants. Among the remaining participants, baseline urine samples were analyzed for participants who had sufficient urine samples for analysis and who were free of diabetes and cardiovascular disease at baseline. After removing missing values in all covariates, 2,925 participants were included in this analysis.

Demographic and lifestyle assessment: Baseline sociodemographic, lifestyle and anthropometric information were obtained through interview and physical examination. The standardized in-person questionnaire included sociodemographic data (age, sex, BMI) and smoking status (never, current, former). Estimated glomerular filtration rate (eGFR), which measures kidney function and might be associated with urinary metal concentrations, was estimated from recalibrated plasma creatinine, age, and sex using the Chronic Kidney Disease – Epidemiology Collaboration formula (19).

Arsenic and cadmium measurements: Metal concentrations were determined using spot urine samples collected at the SHS baseline visit (1989-1991). Metals were measured by inductively coupled plasma mass spectrometry (ICP-MS) at the Trace Element Laboratory of University of Graz, Austria. Inorganic arsenic, monomethylarsonate, dimethylarsinate and arsenobetaine and other arsenic cations were also measured by coupling high performance liquid chromatography (HPLC) and ICP-MS. Arsenic exposure refers to the sum of inorganic and methylated As species. Metal concentrations were corrected by creatinine and log-transformed for analyses.

Cardiovascular disease incidence follow-up: The endpoints of this study are fatal and non-fatal CVD, which were assessed during the follow-up by annual mortality and morbidity surveillance of medical records. Medical records were reviewed and a central adjudication system with two or more physicians was used to classify any potential cardiovascular event (20). Mortality surveillance examined death certificates from state health departments, records from the Indian Health Service, autopsy and coroner's reports, and interviews with physicians or family members. Incident CVD was defined as the first occurrence of fatal or nonfatal coronary heart disease, stroke or congestive heart failure, or other nonfatal CVD. CVD mortality was defined as any fatal CVD. The follow-up time was calculated as the time from urinary sample collection (1989-1991) to the time for CVD events (through 2009, as metal exposure in the tribes changed after the implementation of the EPA Final Arsenic Rule (21)). For participants who did not develop incident CVD, follow-up was censored at the time of occurrence of non-CVD death, loss to follow-up, or the last day of follow-up.

Blood pressure measurements: Participants fasted for 12 hours or more before the physical examination. Brachial artery blood pressure (first and fifth Korotkoff sounds) was measured three consecutive times on seated participants after they had rested 5 minutes with the use of a mercury sphygmomanometer (WA Baum Co) (22). An appropriately sized cuff was placed on the right arm, pulse occlusion pressure was determined, and the cuff was inflated to 20 mm Hg above that pressure. The mean of the last two of these measurements was used for estimation of blood pressure. Systolic blood pressure was measured at three time points: visit 1 (1989-1991), visit 2 (1993-1995) and visit 3 (1998-1999).

*Statistical Analysis*

We used MICE imputation (23) to impute 605 missing values in systolic blood pressure in the second visit and 890 missing values in the third visit, based on the values of the other visits. We ran the mediational g-formula algorithm to evaluate the potential mediating role of blood pressure (measured in three visits) on the association between both cadmium and arsenic (in separate models) and incident CVD, accounting for competing risks by death introducing death as another mediator. To do so, we created a death indicator for each visit (between baseline and the second visit, between the second and the third visit, and between the third visit and the end of follow-up), which we used as the outcome to fit parametric logistic models. From those models, we obtained the probability of dying by that visit using random sampling from the binomial distribution. All models were adjusted for the metal of interest, age, sex, smoking status, BMI and estimated glomerular filtration rate. Thus, we had two sets of mediator models:

1. Models of the mediator of interest (blood pressure) in each visit. These models were adjusted for blood pressure in the previous visit.
2. Models of the competing risk indicator (death) in each visit. These models were adjusted for blood pressure in the same visit.

To fit the outcome model, we used the counting process format (24), in which each individual contributes several rows to the database (one row per time interval, until the event happens). This format can be used in survival analysis without additional specifications and appropriately adjusts for time-varying confounders (25). We used a semi-parametric additive hazards model with CVD incidence as the outcome, adjusted for the metal of interest, blood pressure (in all visits), age, sex, smoking status, BMI and estimated glomerular filtration rate.

We ran the mediational g-formula algorithm with 1000 bootstrap iterations, comparing the metals' 75$^{th}$ percentile to their 25$^{th}$ percentile. We thus have four effects of interest:

- *Total effect*: number of CVD cases per 100,000 person-years, or probability of not developing CVD after 20 years, attributable to an IQR change in each metal concentration.
- *Direct effect*: number of CVD cases per 100,000 person-years, or probability of not developing CVD after 20 years, attributable to an IQR change in each metal concentration not operating through the blood pressure or death pathways.
- *Indirect effect through blood pressure* (the mediator of interest): number of CVD cases per 100,000 person-years, or probability of not developing CVD after 20 years, attributable to an IQR change in each metal concentration that occur through the blood pressure trajectory either causing CVD or causing death (which would make it impossible for CVD to occur). Please note that this pathway can operate through death only through changes in the mediator of interest, not through direct changes in death resulting from changes in the exposure.
- *Indirect effect through death* (the competing event): number of CVD cases per 100,000 person-years, or probability of not developing CVD after 20 years, attributable to an IQR change in each metal concentration avoided by the fact that the individual died before the CVD event happening. Please note that this pathway can operate through death only through changes in the exposure, not through changes in the mediator of interest.

For comparative purposes, we are also interested in the following effects:

- *Sum of the direct effect and the indirect effect through blood pressure*: number of CVD cases per 100,000 person-years, or probability of not developing CVD after 20 years, attributable to an IQR change in each metal concentration that occur through the blood pressure trajectory fixing the survival status to the value it would take under the reference level of the exposure (25$^{th}$ percentile).

- *Direct, indirect and total effects conditional on survival*: effects that are traditionally calculated in mediation analysis, not accounting for competing risks.

*Results*

Participant characteristics are shown in Table. There were 977 CVD cases. Participants who had a CVD event were older, more likely to be current smokers, had higher SBP levels and both urinary arsenic and cadmium concentrations. Table 2 shows the results of the longitudinal mediation analysis for arsenic, both without accounting for competing risks (i.e., not taking into account the pathway through death for other causes different to CVD) and accounting for competing risks. Results are presented both in differences in hazards and survival probability differences. Of 345 CVD cases attributable to an interquartile range increase in log-arsenic exposure, 45 would happen through the blood pressure trajectory, and 26 CVD cases would be avoided by the participant dying from other causes before having the chance to develop CVD per 100,000 person-years. For the difference in hazards, the indirect effect through blood pressure is statistically significant when not accounting for competing risks, but it is not when accounting for competing risks, as the confidence intervals become much wider. It is, however, statistically significant in the survival probability difference scale. The total effect is statistically significant but attenuated when accounting for competing risks, as this estimator takes into account the CVD cases that are not allowed to happen given that the individual dies before, and is not, therefore, in the risk set.

Table 3 shows the results for cadmium. Of 137 CVD cases attributable to a interquartile range increase in log-cadmium exposure, of which, 14 would happen through the blood pressure trajectory, and 22 CVD cases would be avoided by the participant dying before having the chance to develop CVD. We see no statistically significant indirect effect in the hazard difference scale, regardless of accounting or not for competing risks. For the survival probability difference scale, the indirect effect is not statistically significant after accounting for competing risks. The total effect is not statistically significant and is attenuated when accounting for competing risks. The indirect effect through death has an impact on the total effect but is not statistically significant for either cadmium or arsenic.

## 8. Discussion

In this work, we developed an extension of the mediational g-formula which can deal with competing events. This is, to our knowledge, the first method that incorporates competing events to the mediation analysis setting using the path-specific effects framework. Our algorithm reports effects in terms of attributable cases per a number of person-years, which is a measure of public health impact, and also in the survival probability difference scale, which has a causal interpretation.

Traditional Cox proportional hazards models censor people that die as if they had dropped out of the study, and therefore consider they could develop the disease at the same rate as those who remain in the study, leading to potential biases of the associations. Other approaches for competing risks in mediation analysis include the survivor average causal effect (26) and the separable effects framework (27) These methods, however, are focused on a population that is either unrealistic (all survivors), or not observed (those who would survive no matter the exposure level). Also, effects cannot be separable in all settings. For example, in this setting, we have intertwined mediators, thus, we would not be able to consider separate non-intertwined paths for the mediator of interest and the competing event. Instead, the indirect effect through the mediator of interest can also operate through the death pathway, provided that the death trajectory changes because of changes in the mediator of interest, and not because of direct changes in the exposure. Our method decomposes the contribution of each pathway to the association between an exposure and a health outcome while accounting for people that die during follow-up. This provides a better opportunity to investigate different pathways involved in the adverse health effects of elevated urinary metal levels.

Both exposure-mediator interactions and non-linear effects can be considered in our algorithm. The algorithm can be extended to include time-varying confounders by considering an additional path-specific effect that evaluates the effect of the exposure on the outcome through those time-varying confounders, nested with the path through the mediator of interest and through the competing event. In addition, we currently model the trajectory of the mediator of interest by adjusting the mediator models at each time point for the immediate previous time point of the mediator, i.e., we do not adjust for all the previous mediator measurements to avoid multi-collinearity. This modelling approach could potentially be modified to include more specific measures of the longitudinal trajectory of the mediator such as the rate of change.

Using our mediational g-formula algorithm, we identified a statistically significant indirect effect of the trajectory of blood pressure over time on the association between arsenic and CVD, but not cadmium, and CVD. Our results also show the importance of taking competing risks into account for cadmium, as many people typically die from other causes attributable to cadmium before being able to develop CVD. The wider confidence intervals in both the indirect and total effects show that the statistical uncertainty is greater when accounting for competing risks. The total effect not accounting for competing risks does not consider any variation in the survival status, as it is conditional on survival. Conversely, in our algorithm, we account for the difference in the survival trajectory under the two exposure scenarios, which makes this approach likely to enable modeling of a more realistic scenario. The effects were more statistically significant when using the survival probability difference in 20 years, which has a causal interpretation, as compared to using differences in hazards.

This work has several limitations. First, we consider discrete time points for death indicators, rather than modeling death as time-to-event, which might potentially lead to some measurement errors due to not being able to incorporate the exact time in which the event happened. An extension to multi-state models (28), in which our primary event of interest (CVD, in this case) and death could be considered as a composite event, represents relevant future work. Another limitation is the parametric modeling approach. It would be of interest to extend this work to the machine learning setting, to be able to consider several exposures simultaneously, for example. In addition, the no unmeasured confounding assumption is not verifiable in practice in observational studies (29). Thus, the results of this study need to be interpreted with caution. The development of sensitivity analyses that evaluate how strong an unmeasured confounder would need to be to bias the estimated effects is left for future work.

In conclusion, this work shows the importance of considering competing events in mediation analysis with survival outcomes. The fact of being able to identify the effects through different paths sheds light on the impact of metals on the outcome of interest, as well as through competing risks, more transparently.

**Table 1.** Participant characteristics by cardiovascular disease status.

| | Overall (N=2,932) | Non-cases (N=1,955) | Cardiovascular disease cases (N=977) |
|---|---|---|---|
| Age (years) | 55.3 (49.3, 62.6) | 54.1 (48.6, 61.4) | 57.4 (51.2, 64.5) |
| Sex (%) | | | |
| Female | 1698 (57.9) | 1151 (58.9) | 547 (56) |
| Male | 1234 (42.1) | 804 (41.1) | 430 (44) |
| Study center (%) | | | |
| AZ | 397 (13.5) | 294 (15) | 103 (10.5) |
| OK | 1275 (43.5) | 891 (45.6) | 384 (39.3) |
| ND/SD | 1260 (43) | 770 (39.4) | 490 (50.2) |
| Smoking (%) | | | |
| Never | 851 (29) | 593 (30.3) | 258 (26.4) |
| Former | 979 (33.4) | 656 (33.6) | 323 (33.1) |
| Current | 1102 (37.6) | 706 (36.1) | 396 (40.5) |
| BMI (kg/m2) | 29.71 (26.38, 33.77) | 29.38 (25.9, 33.46) | 30.25 (27.12, 34.23) |
| eGFR (mL/min/1.73m$^2$) | 99.98 (90.58, 107.3) | 100.72 (91.64, 107.89) | 98.2 (89.19, 105.79) |
| SBP (visit 1) | 124 (113, 136.12) | 122 (111, 134) | 128 (118, 141) |
| SBP (visit 2) | 126 (114, 140) | 124 (113, 136) | 131 (119, 144) |
| SBP (visit 3) | 129 (118, 142) | 128 (117, 141) | 132 (120, 146) |
| Arsenic µg/g | 8.41 (5.12, 14.27) | 8.36 (4.99, 13.9) | 8.56 (5.36, 14.78) |
| Cadmium µg/g | 0.97 (0.62, 1.5) | 0.95 (0.61, 1.47) | 1 (0.64, 1.55) |

Median (interquartile range) for continuous variables, percentages for categorical variables

**Table 2.** Direct, indirect, and total effects of one interquartile range increase in urinary arsenic on incident CVD through the systolic blood pressure trajectory and death.

| | Without accounting for competing risks | | Accounting for competing risks | |
|---|---|---|---|---|
| | Hazard difference[a] | Difference in survival probability in 20 years (%) | Hazard difference[a] | Difference in survival probability in 20 years (%) |
| Direct effect | 330.6 (39.7, 617.9) | -2.6 (-2.6, -2.6) | 330.6 (36.2, 627.4) | -2.59 (-2.60, -2.58) |
| Indirect effect through the blood pressure trajectory | 44.9 (9.6, 86.3) | -0.4 (-0.4, -0.3) | 46.5 (7.5, 90.4) | -0.34 (-0.70, -0.02) |
| Indirect effect through the death process | - | - | -8.9 (-22.5, 4.5) | 0.22 (-0.07, 0.49) |
| Total effect | 375.8 (85.9, 666.1) | -2.91 (-2.93, -2.90) | 368.6 (72.1, 668.4) | -2.73 (-2.99, -2.36) |
| Sum of direct effect and indirect effect through blood pressure | 375.6 (49.3, 704.2) | -2.91 (-2.93, -2.89) | 377.1 (43.7, 717.8) | -2.93 (-3.30, -2.60) |

[a] The hazard difference is reported in number of cases per 100,000 person-years.
Models adjusted for age, sex, study center (Arizona, Oklahoma, or North Dakota and South Dakota), estimated glomerular filtration rate, BMI and smoking status (never, former, or current smoking).

**Table 3.** Direct, indirect, and total effects of one interquartile range increase in urinary cadmium on incident CVD through the systolic blood pressure trajectory and death.

| | Without accounting for competing risks | | Accounting for competing risks | |
|---|---|---|---|---|
| | Hazard difference[a] | Difference in survival probability in 20 years (%) | Hazard difference[a] | Difference in survival probability in 20 years (%) |
| Direct effect | 149.4 (-80.7, 374.4) | -1.68 (-1.70, -1.67) | 146.2 (-83.2, 374.9) | -1.69 (-1.71, -1.67) |
| Indirect effect through the blood pressure trajectory | 13.1 (-17.7, 46.1) | -0.152 (-0.153, -0.150) | 13.8 (-19.8, 49.1) | -0.16 (-0.32, -0.02) |
| Indirect effect through the death process | - | - | -8.8 (-22.6, 4.7) | 0.1 (-0.04, 0.23) |
| Total effect | 161.8 (-72.1, 387.8) | -1.83 (-1.85, -1.81) | 152.2 (-80.2, 382.9) | -1.76 (-1.9, -1.63) |
| Sum of direct effect and indirect effect through blood pressure | 162.5 (-98.4, 420.5) | -1.83 (-1.85, -1.82) | 160.0 (-103.0, 424.0) | -1.85 (-2.03, -1.69) |

[a] The hazard difference is reported in number of cases per 100,000 person-years.
Models adjusted for age, sex, study center (Arizona, Oklahoma, or North Dakota and South Dakota), estimated glomerular filtration rate, BMI and smoking status (never, former, or current smoking).

**SUPPLEMENTARY MATERIAL**

**A path-specific effect approach to mediation analysis with time-varying mediators and time-to-event outcomes accounting for competing risks**

**Supplementary Figure 1.** Directed acyclic graph for the direct effect of the exposure on the outcome.

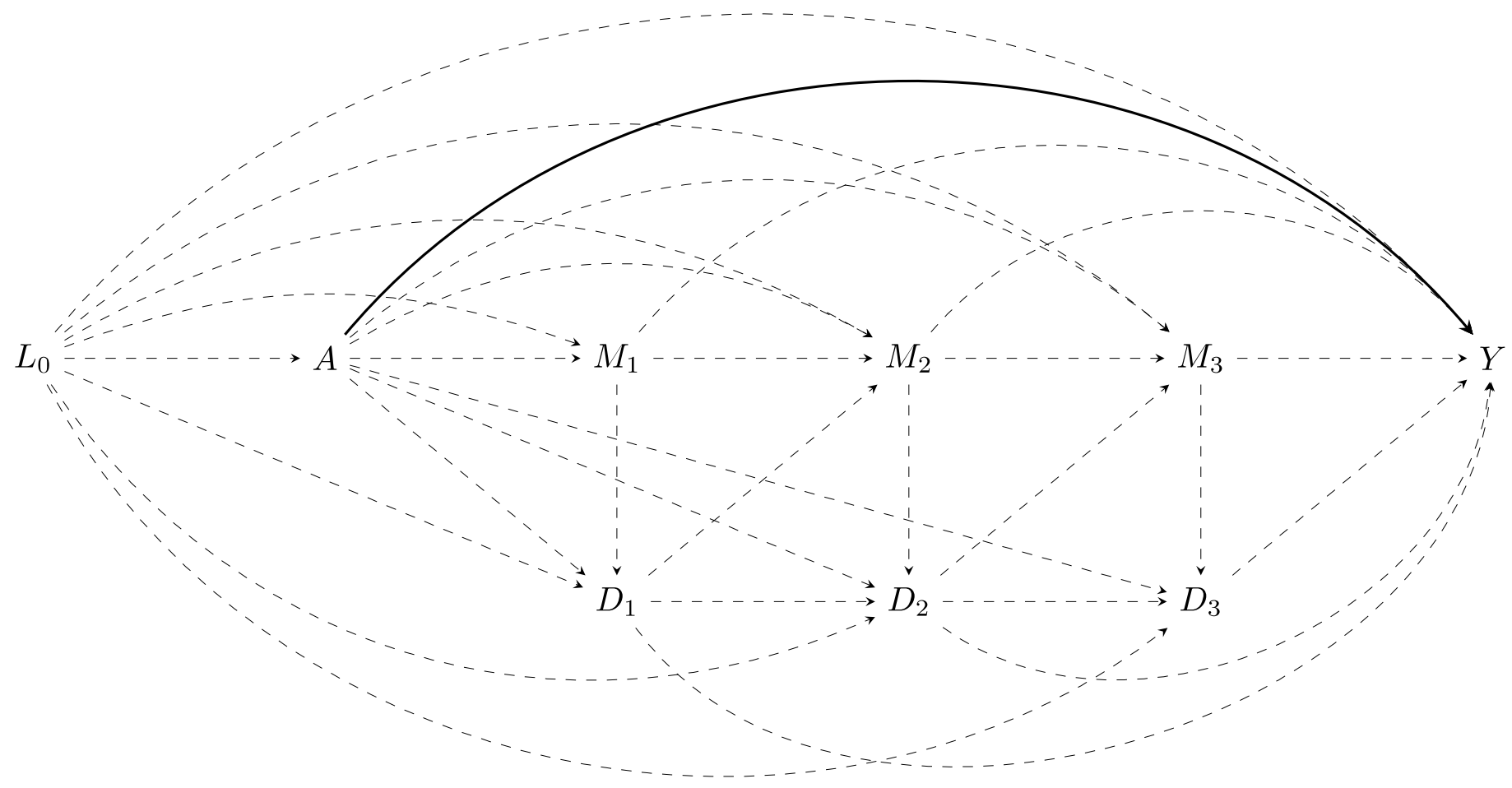

**Supplementary Figure 2.** Directed acyclic graph for the indirect effect through the mediator of interest.

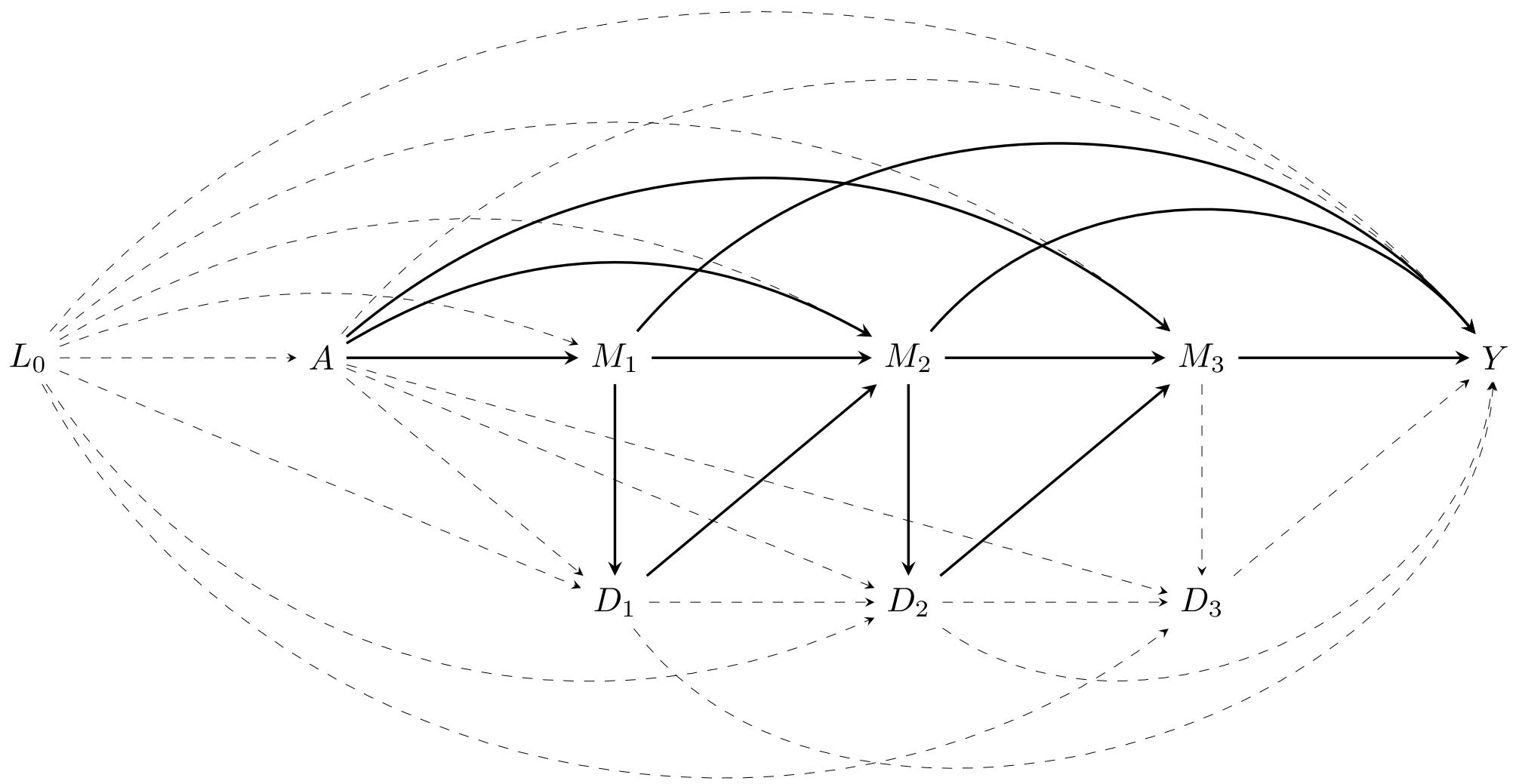

**Supplementary Figure 3.** Directed acyclic graph for the indirect effect through death/competing event.

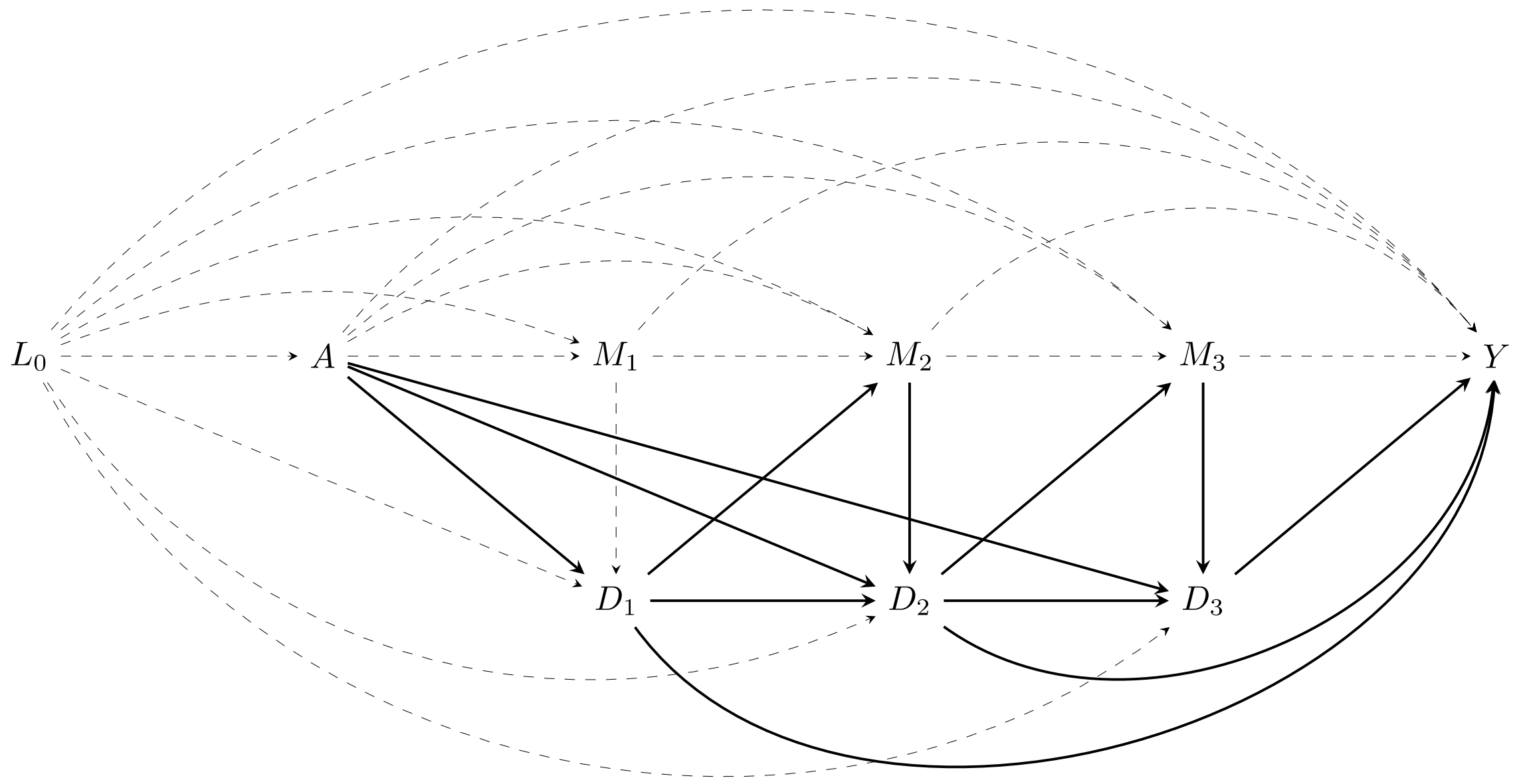

**Supplementary Figure 4.** Directed acyclic graph for the sum of the direct effect and the indirect effect through the mediator of interest

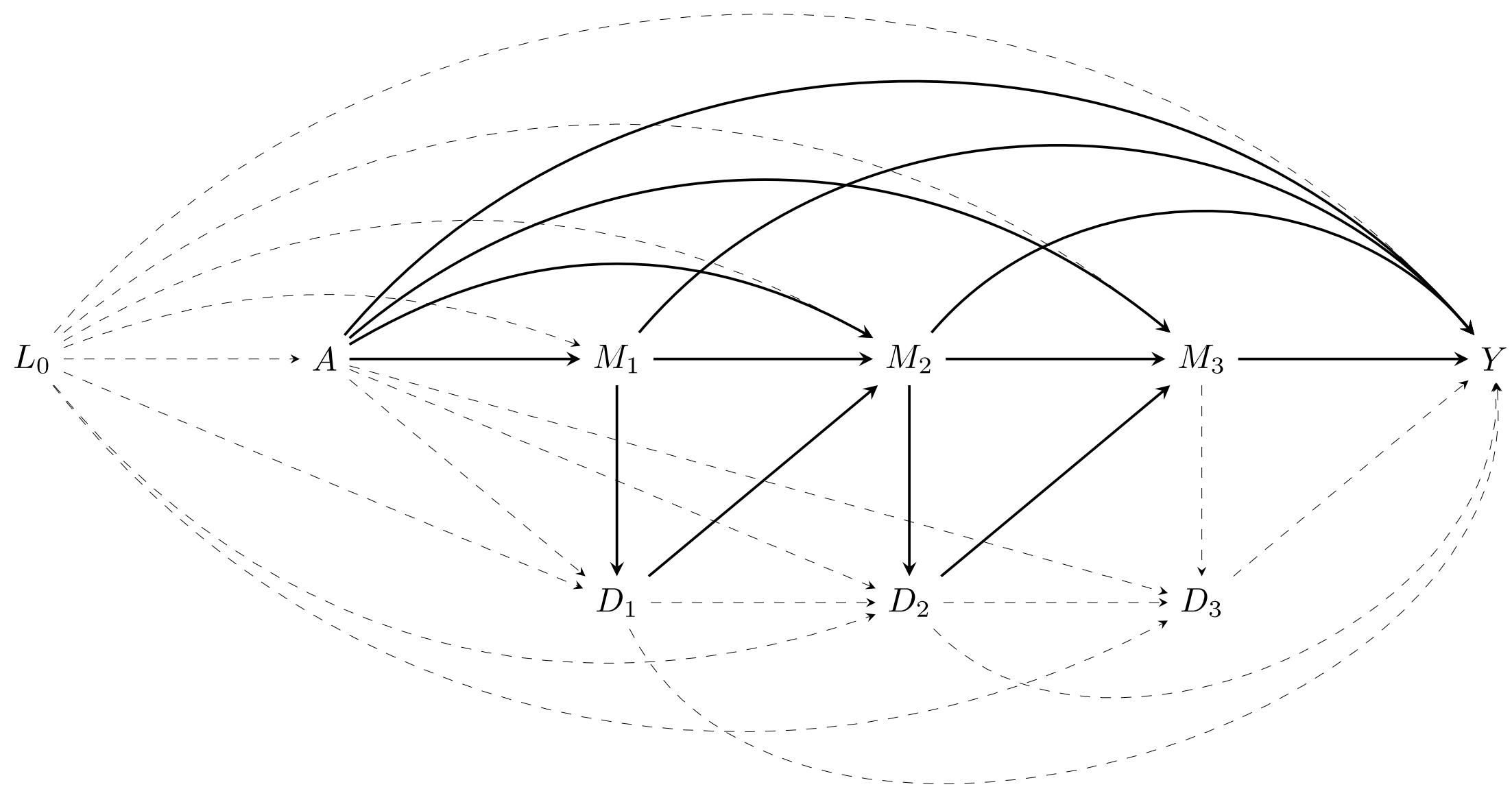

## Supplementary Methods

*Proof of the calculation of the total effect using the mediational g-formula.*

$$\phi(a^*,a^*,a^*) \equiv \mathrm{E}\left(Y_{k+1}\left(a^*,\bar{M}_k\left(a^*,\bar{S}_{k-1}(a^*)\right),\bar{S}_k\left(a^*,\bar{M}_k(a^*)\right)\right)\middle|L_0=l_0\right)$$

$$= \mathrm{E}\left(Y_{k+1}\left(a^*,\bar{M}_k\left(a^*,\bar{S}_{k-1}(a^*)\right),\bar{S}_k\left(a^*,\bar{M}_k(a^*)\right)\right)\middle|\bar{S}_k\left(a^*,\bar{M}_k(a^*)\right)=1,L_0=l_0\right)P\left(\bar{S}_k\left(a^*,\bar{M}_k(a^*)\right)=1\middle|L_0=l_0\right)$$

$$+ \mathrm{E}\left(Y_{k+1}\left(a^*,\bar{M}_k\left(a^*,\bar{S}_{k-1}(a^*)\right),\bar{S}_k\left(a^*,\bar{M}_k(a^*)\right)\right)\middle|\bar{S}_k\left(a^*,\bar{M}_k(a^*)\right)=0,L_0=l_0\right)P\left(\bar{S}_k\left(a^*,\bar{M}_k(a^*)\right)=0\middle|L_0=l_0\right)$$

$\because$ **theorem of the total probability**

$$= \mathrm{E}\left(Y_{k+1}\left(a^*,\bar{s}_k=1,\bar{M}_k(a^*,\bar{s}_{k-1}=1)\right)\middle|\bar{S}_k\left(a^*,\bar{M}_k(a^*)\right)=1,L_0=l_0\right)P\left(\bar{S}_k\left(a^*,\bar{M}_k(a^*)\right)=1\middle|L_0=l_0\right)$$

$\left(\because \boldsymbol{Y\left(\bar{S}_k\left(a^*,\bar{M}_k(a^*)\right)=0\right)=0}\right)$

$$= \mathrm{E}\left(\bar{S}_k\left(a^*,\bar{M}_k(a^*)\right)Y_{k+1}\left(a^*,\bar{s}_k=1,\bar{M}_k(a^*,\bar{s}_{k-1}=1)\right)\middle|L_0=l_0\right)$$

$\left(\because \textbf{B is binary} \Rightarrow \boldsymbol{E(BY)=E(Y|B=1)P(B=1)}\right)$

$$= \int_{\bar{m}_k} \mathrm{E}(\bar{S}_k(a^*,\bar{m}_k)\,Y_{k+1}(a^*,1,\bar{m}_k)|\bar{M}_k(a^*,\bar{s}_{k-1}=1)=\bar{m}_k,L_0=l_0)f(\bar{M}_k(a^*,\bar{s}_{k-1}=1)=\bar{m}_k|\bar{S}_{k-1}=1,L_0=l_0)\,d\bar{m}_k$$

$$= \int_{\bar{m}_k} \mathrm{E}(\bar{S}_k(a^*,\bar{m}_k)\,Y_{k+1}(a^*,1,\bar{m}_k)|\bar{M}_k(a^*,\bar{s}_{k-1}=1)=\bar{m}_k,A=a^*,L_0=l_0)f(\bar{M}_k(a^*,1)=\bar{m}_k|\bar{S}_{k-1}=1,L_0=l_0)d\bar{m}_k$$

$\because \boldsymbol{Y_{k+1}(a^*,\bar{s}_k=1,\bar{m}_k)\perp A\,|L_0}$ **and** $\boldsymbol{\bar{S}_k(a^*,\bar{m}_k)\perp A\,|L_0}$

$$= \int_{\bar{m}_k} \mathrm{E}(\bar{S}_k(a^*,\bar{m}_k)\,Y_{k+1}(a^*,1,\bar{m}_k)|A=a^*,L_0=l_0)f(\bar{M}_k(a^*,1)=\bar{m}_k|\bar{S}_{k-1}=1,L_0=l_0)\,d\bar{m}_k$$

$\because \boldsymbol{\left(Y_{k+1}(a^*,\bar{s}_k=1,\bar{m}_k),\bar{S}_k(a^*,\bar{m}_k)\right)\perp\bar{M}_k\,|A,L_0}$

$$= \int_{\bar{m}_k} \mathrm{E}(\bar{S}_k(a^*,\bar{m}_k)\,Y_{k+1}(a^*,1,\bar{m}_k)|A=a^*,\bar{M}_k=\bar{m}_k,L_0=l_0)\,f(\bar{M}_k(a^*,1)=\bar{m}_k|\bar{S}_{k-1}=1,L_0=l_0)\,d\bar{m}_k$$

$\because \boldsymbol{Y_{k+1}(a^*,\bar{s}_k=1,\bar{m}_k)\perp\bar{M}_k|A,\bar{S}_{k-1}=1,L_0}$ **and** $\boldsymbol{\bar{S}_k(a^*,\bar{m}_k)\perp\bar{M}_k|A,L_0}$

$$= \int_{\bar{m}_k} \mathrm{E}(\bar{S}_k Y_{k+1}(a^*, 1, \bar{m}_k) | A = a^*, \bar{M}_k = \bar{m}_k, L_0 = l_0) f(\bar{M}_k(a^*, 1)$$

$$= \bar{m}_k | \bar{S}_{k-1} = 1, L_0 = l_0) d\bar{m}_k$$

$\because$ ***consistency***

$$= \int_{\bar{m}_k} \mathrm{E}(Y_{k+1}(a^*, 1, \bar{m}_k) | \bar{S}_k = 1, A = a^*, \bar{M}_k = \bar{m}_k, L_0 = l_0) f(\bar{S}_k$$

$$= 1 | A = a^*, \bar{M}_k = \bar{m}_k, C = c) f(\bar{M}_k(a^*, 1) = \bar{m}_k | \bar{S}_{k-1} = 1, L_0 = l_0) d\bar{m}_k$$

$$= \int_{\bar{m}_k} \mathrm{E}(Y_{k+1}(a^*, 1, \bar{m}_k) | \bar{S}_k = 1, A = a^*, \bar{M}_k = \bar{m}_k, L_0 = l_0)\ f(\bar{S}_k$$

$$= 1 |\ A = a^*, \bar{M}_k = \bar{m}_k, L_0 = l_0)\ f(\bar{M}_k(a^*, 1)$$

$$= \bar{m}_k |\ A = a^*, \bar{S}_{k-1} = 1, L_0 = l_0)\ d\bar{m}_k$$

$\because \boldsymbol{\bar{M}_k(a^*, \bar{s}_{k-1} = 1) \perp A\ | \bar{S}_{k-1} = 1, L_0}$

$$= \int_{\bar{m}_k} \mathrm{E}(Y_{k+1} | A = a^*, \bar{S}_k = 1, \bar{M}_k = \bar{m}_k, L_0 = l_0)\ f(\bar{S}_k = 1 | A = a^*, \bar{M}_k = \bar{m}_k, L_0 = l_0)\ f(\bar{M}_k$$

$$= \bar{m}_k | A = a^*, \bar{S}_{k-1} = 1, L_0 = l_0) d\bar{m}_k$$

$\because$ ***consistency***

On the other hand:

$$\phi(a, a, a) \equiv \mathrm{E}\left(Y_{k+1}\left(a, \bar{M}_k\left(a, \bar{S}_{k-1}(a)\right), \bar{S}_k\left(a, \bar{M}_k(a)\right)\right) \Big| L_0 = l_0\right)$$

$$= \mathrm{E}\left(Y_{k+1}\left(a, \bar{M}_k\left(a, \bar{S}_{k-1}(a)\right), \bar{S}_k\left(a, \bar{M}_k(a)\right)\right) \Big| \bar{S}_k\left(a, \bar{M}_k(a)\right) = 1, L_0 = l_0\right) P\left(\bar{S}_k\left(a, \bar{M}_k(a)\right)\right.$$

$$\left. = 1 | L_0 = l_0\right)$$

$$+\ \mathrm{E}\left(Y_{k+1}\left(a, \bar{M}_k\left(a, \bar{S}_{k-1}(a)\right), \bar{S}_k\left(a, \bar{M}_k(a)\right)\right) \Big| \bar{S}_k\left(a, \bar{M}_k(a)\right) = 0, L_0 = l_0\right) P\left(\bar{S}_k\left(a, \bar{M}_k(a)\right)\right.$$

$$\left. = 0 | L_0 = l_0\right)$$

$\because$ **theorem of the total probability**

$$= \mathrm{E}\left(Y_{k+1}\left(a, \bar{s}_k = 1, \bar{M}_k(a, \bar{s}_{k-1} = 1)\right) \middle| \bar{S}_k\left(a, \bar{M}_k(a)\right) = 1, L_0 = l_0\right) P\left(\bar{S}_k\left(a, \bar{M}_k(a)\right) = 1 | L_0 = l_0\right)$$

$\left(\because \boldsymbol{Y\left(\bar{S}_k\left(a, \bar{M}_k(a)\right) = 0\right) = 0}\right)$

$$= \mathrm{E}\left(\bar{S}_k\left(a, \bar{M}_k(a)\right) Y_{k+1}\left(a, \bar{s}_k = 1, \bar{M}_k(a, \bar{s}_{k-1} = 1)\right) | L_0 = l_0\right)$$

$\left(\because \textbf{Bisbinary} \Rightarrow \boldsymbol{E(BY) = E(Y|B = 1)P(B = 1)}\right)$

$$= \int_{\bar{m}_k} \mathrm{E}(\bar{S}_k(a,\bar{m}_k)\ Y_{k+1}(a,1,\bar{m}_k)|\bar{M}_k(a,\bar{s}_{k-1}=1)=\bar{m}_k, L_0=l_0) f(\bar{M}_k(a,\bar{s}_{k-1}=1)$$

$$= \bar{m}_k|\bar{S}_{k-1}=1, L_0=l_0)\ d\bar{m}_k$$

$$= \int_{\bar{m}_k} \mathrm{E}(\bar{S}_k(a,\bar{m}_k)\ Y_{k+1}(a,1,\bar{m}_k)|\bar{M}_k(a,\bar{s}_{k-1}=1)=\bar{m}_k, A=a, L_0=l_0) f(\bar{M}_k(a,1)$$

$$= \bar{m}_k|\bar{S}_{k-1}=1, L_0=l_0) d\bar{m}_k$$

$\because \boldsymbol{Y_{k+1}(a,\bar{s}_k=1,\bar{m}_k)\perp A\,|L_0}$ **and** $\boldsymbol{\bar{S}_k(a,\bar{m}_k)\perp A\,|L_0}$

$$= \int_{\bar{m}_k} \mathrm{E}(\bar{S}_k(a,\bar{m}_k)\ Y_{k+1}(a,1,\bar{m}_k)|A=a, L_0=l_0) f(\bar{M}_k(a,1)=\bar{m}_k|\bar{S}_{k-1}=1, L_0=l_0)\ d\bar{m}_k$$

$\because \boldsymbol{\left(Y_{k+1}(a,\bar{s}_k=1,\bar{m}_k), \bar{S}_k(a,\bar{m}_k)\right)\perp \bar{M}_k\,|A, L_0}$

$$= \int_{\bar{m}_k} \mathrm{E}(\bar{S}_k(a,\bar{m}_k)\ Y_{k+1}(a,1,\bar{m}_k)|A=a, \bar{M}_k=\bar{m}_k, L_0=l_0) f(\bar{M}_k(a,1)$$

$$= \bar{m}_k|\bar{S}_{k-1}=1, L_0=l_0)\ d\bar{m}_k$$

$\because \boldsymbol{Y_{k+1}(a,\bar{s}_k=1,\bar{m}_k)\perp \bar{M}_k|A,\bar{S}_{k-1}=1, L_0}$ **and** $\boldsymbol{\bar{S}_k(a,\bar{m}_k)\perp \bar{M}_k|A,L_0}$

$$= \int_{\bar{m}_k} \mathrm{E}(\bar{S}_k Y_{k+1}(a,1,\bar{m}_k)|A=a, \bar{M}_k=\bar{m}_k, L_0=l_0) f(\bar{M}_k(a,1)=\bar{m}_k|\bar{S}_{k-1}=1, L_0=l_0) d\bar{m}_k$$

$\because$ ***consistency***

$$= \int_{\bar{m}_k} \mathrm{E}(Y_{k+1}(a,1,\bar{m}_k)|\bar{S}_k=1, A=a, \bar{M}_k=\bar{m}_k, L_0=l_0) f(\bar{S}_k$$

$$= 1|A=a, \bar{M}_k=\bar{m}_k, L_0=l_0) f(\bar{M}_k(a,1)=\bar{m}_k|\bar{S}_{k-1}=1, L_0=l_0) d\bar{m}_k$$

$$= \int_{\bar{m}_k} \mathrm{E}(Y_{k+1}(a,1,\bar{m}_k)|\bar{S}_k=1, A=a, \bar{M}_k=\bar{m}_k, L_0=l_0) f(\bar{S}_k$$

$$= 1|A=a, \bar{M}_k=\bar{m}_k, L_0=l_0) f(\bar{M}_k(a,1)=\bar{m}_k|A=a, \bar{S}_{k-1}=1, L_0=l_0) d\bar{m}_k$$

$\because \boldsymbol{\bar{M}_k(a,\bar{s}_{k-1}=1)\perp A\,|\bar{S}_{k-1}=1, L_0}$

$$= \int_{\bar{m}_k} \mathrm{E}(Y_{k+1}|A=a, \bar{S}_k=1, \bar{M}_k=\bar{m}_k, L_0=l_0) f(\bar{S}_k=1|A=a, \bar{M}_k=\bar{m}_k, L_0=l_0)\ f(\bar{M}_k$$

$$= \bar{m}_k|A=a, \bar{S}_{k-1}=1, L_0=l_0) d\bar{m}_k$$

$\because$ ***consistency***

*Proof of the calculation of the indirect through the mediator of interest using the mediational g-formula.*

$\phi(a^*, a^*, a) \equiv \mathrm{E}\left(Y_{k+1}\left(a^*, \bar{M}_k\left(a, \bar{S}_{k-1}(a^*)\right), \bar{S}_k\left(a^*, \bar{M}_k(a)\right)\right) \middle| L_0 = l_0\right)$

$= \mathrm{E}\left(Y_{k+1}\left(a^*, \bar{M}_k\left(a, \bar{S}_{k-1}(a^*)\right), \bar{S}_k\left(a^*, \bar{M}_k(a)\right)\right) \middle| \bar{S}_k\left(a^*, \bar{M}_k(a)\right) = 1, L_0 = l_0\right) P\left(\bar{S}_k\left(a^*, \bar{M}_k(a)\right)\right.$
$\left. = 1 | L_0 = l_0\right)$

$+ \ \mathrm{E}\left(Y_{k+1}\left(a^*, \bar{M}_k\left(a, \bar{S}_{k-1}(a^*)\right), \bar{S}_k\left(a^*, \bar{M}_k(a)\right)\right) \middle| \bar{S}_k\left(a^*, \bar{M}_k(a)\right) = 0, L_0 = l_0\right) P\left(\bar{S}_k\left(a^*, \bar{M}_k(a)\right)\right.$
$\left. = 0 | L_0 = l_0\right)$

$\because$ **theorem of the total probability**

$= \mathrm{E}\left(Y_{k+1}\left(a^*, \bar{s}_k = 1, \bar{M}_k(a, \bar{s}_{k-1} = 1)\right) \middle| \bar{S}_k\left(a^*, \bar{M}_k(a)\right) = 1, L_0 = l_0\right) P\left(\bar{S}_k\left(a^*, \bar{M}_k(a)\right) =\right.$
$\left. 1 | L_0 = l_0\right)$

$\left(\because \boldsymbol{Y\left(\bar{S}_k\left(a^*, \bar{M}_k(a)\right) = 0\right) = 0}\right)$

$= \mathrm{E}\left(\bar{S}_k\left(a^*, \bar{M}_k(a)\right) Y_{k+1}\left(a^*, \bar{s}_k = 1, \bar{M}_k(a, \bar{s}_{k-1} = 1)\right) | L_0 = l_0\right)$

$\left(\because \textbf{Bisbinary} \Rightarrow \boldsymbol{E(BY) = E(Y|B=1)P(B=1)}\right)$

$$= \int_{\bar{m}_k} \mathrm{E}(\bar{S}_k(a^*, \bar{m}_k)\, Y_{k+1}(a^*, 1, \bar{m}_k) | \bar{M}_k(a, \bar{s}_{k-1} = 1) = \bar{m}_k, L_0 = l_0) f(\bar{M}_k(a, \bar{s}_{k-1} = 1) = \bar{m}_k | \bar{S}_{k-1} = 1, L_0 = l_0)\, d\bar{m}_k$$

$$= \int_{\bar{m}_k} \mathrm{E}(\bar{S}_k(a^*, \bar{m}_k)\, Y_{k+1}(a^*, 1, \bar{m}_k) | \bar{M}_k(a, \bar{s}_{k-1} = 1) = \bar{m}_k, A = a^*, L_0 = l_0) f(\bar{M}_k(a, \bar{s}_{k-1} = 1) = \bar{m}_k | \bar{S}_{k-1} = 1, L_0 = l_0)\, d\bar{m}_k$$

$\because \boldsymbol{Y_{k+1}(a^*, \bar{s}_k = 1, \bar{m}_k) \perp A | L_0}$ **and** $\boldsymbol{\bar{S}_k(a^*, \bar{m}_k) \perp A | L_0}$

$$= \int_{\bar{m}_k} \mathrm{E}(\bar{S}_k(a^*, \bar{m}_k)\, Y_{k+1}(a^*, 1, \bar{m}_k) | A = a^*, L_0 = l_0) f(\bar{M}_k(a, 1) = \bar{m}_k | \bar{S}_{k-1} = 1, L_0 = l_0)\, d\bar{m}_k$$

$\because \boldsymbol{\left(Y_{k+1}(a^*, \bar{s}_k = 1, \bar{m}_k), \bar{S}_k(a^*, \bar{m}_k)\right) \perp \bar{M}_k(a, \bar{s}_{k-1} = 1) | A, L_0}$

$$= \int_{\bar{m}_k} \mathrm{E}(\bar{S}_k(a^*, \bar{m}_k)\, Y_{k+1}(a^*, 1, \bar{m}_k) | A = a^*, \bar{M}_k = \bar{m}_k, L_0 = l_0) f(\bar{M}_k(a, 1) = \bar{m}_k | \bar{S}_{k-1} = 1, L_0 = l_0) d\bar{m}_k$$

$\because \boldsymbol{Y_{k+1}(a^*, \bar{s}_k = 1, \bar{m}_k) \perp \bar{M}_k | A, \bar{S}_{k-1} = 1, L_0}$ **and** $\boldsymbol{\bar{S}_k(a^*, \bar{m}_k) \perp \bar{M}_k | A, L_0}$

$$= \int_{\bar{m}_k} \mathrm{E}(\bar{S}_k Y_{k+1}(a^*, 1, \bar{m}_k)|A = a^*, \bar{M}_k = \bar{m}_k, L_0 = l_0) f(\bar{M}_k(a, 1)$$

$$= \bar{m}_k | \bar{S}_{k-1} = 1, L_0 = l_0) d\bar{m}_k$$

$\because$ ***consistency***

$$= \int_{\bar{m}_k} \mathrm{E}(Y_{k+1}(a^*, 1, \bar{m}_k)|\bar{S}_k = 1, A = a^*, \bar{M}_k = \bar{m}_k, L_0 = l_0) f(\bar{S}_k$$

$$= 1|A = a^*, \bar{M}_k = \bar{m}_k, L_0 = l_0) f(\bar{M}_k(a, 1) = \bar{m}_k | \bar{S}_{k-1} = 1, L_0 = l_0) d\bar{m}_k$$

$$= \int_{\bar{m}_k} \mathrm{E}(Y_{k+1}(a^*, 1, \bar{m}_k)|\bar{S}_k = 1, A = a^*, \bar{M}_k = \bar{m}_k, L_0 = l_0) f(\bar{S}_k$$

$$= 1|A = a^*, \bar{M}_k = \bar{m}_k, L_0 = l_0) f(\bar{M}_k(a, 1) = \bar{m}_k | A = a, \bar{S}_{k-1} = 1, L_0 = l_0) d\bar{m}_k$$

$\because \boldsymbol{\bar{M}_k(a, \bar{s}_{k-1} = 1) \perp A \,|\bar{S}_{k-1} = 1, L_0}$

$$= \int_{\bar{m}_k} \mathrm{E}(Y_{k+1}|A = a^*, \bar{S}_k = 1, \bar{M}_k = \bar{m}_k, L_0 = l_0) f(\bar{S}_k = 1|A = a^*, \bar{M}_k = \bar{m}_k, L_0 = l_0) f(\bar{M}_k$$

$$= \bar{m}_k | A = a, \bar{S}_{k-1} = 1, L_0 = l_0) d\bar{m}_k$$

$\because$ ***consistency***

On the other hand:

$$\phi(a^*, a^*, a^*) \equiv \mathrm{E}\left(Y_{k+1}\left(a^*, \bar{M}_k\big(a^*, \bar{S}_{k-1}(a^*)\big), \bar{S}_k\big(a^*, \bar{M}_k(a^*)\big)\right) \middle| L_0 = l_0\right)$$

$$= \mathrm{E}\left(Y_{k+1}\left(a^*, \bar{M}_k\big(a^*, \bar{S}_{k-1}(a^*)\big), \bar{S}_k\big(a^*, \bar{M}_k(a^*)\big)\right) \middle| \bar{S}_k\big(a^*, \bar{M}_k(a^*)\big) = 1, L_0 = l_0\right) P\big(\bar{S}_k\big(a^*, \bar{M}_k(a^*)\big)$$

$$= 1|L_0 = l_0\big)$$

$$+ \ \mathrm{E}\left(Y_{k+1}\left(a^*, \bar{M}_k\big(a^*, \bar{S}_{k-1}(a^*)\big), \bar{S}_k\big(a^*, \bar{M}_k(a^*)\big)\right) \middle| \bar{S}_k\big(a^*, \bar{M}_k(a^*)\big) = 0, L_0 = l_0\right) P\big(\bar{S}_k\big(a^*, \bar{M}_k(a^*)\big)$$

$$= 0|L_0 = l_0\big)$$

$\because$ **theorem of the total probability**

$$= \mathrm{E}\big(Y_{k+1}\big(a^*, \bar{s}_k = 1, \bar{M}_k(a^*, \bar{s}_{k-1} = 1)\big)\big|\bar{S}_k\big(a^*, \bar{M}_k(a^*)\big) = 1, L_0 = l_0\big) P\big(\bar{S}_k\big(a^*, \bar{M}_k(a^*)\big) =$$

$$1|L_0 = l_0\big)$$

$\big(\because \boldsymbol{Y\big(\bar{S}_k\big(a^*, \bar{M}_k(a)\big) = 0\big) = 0}\big)$

$$= \mathrm{E}\big(\bar{S}_k\big(a^*, \bar{M}_k(a^*)\big)\, Y_{k+1}\big(a^*, \bar{s}_k = 1, \bar{M}_k(a^*, \bar{s}_{k-1} = 1)\big)|L_0 = l_0\big)$$

$\big(\because$ **Bisbinary** $\Rightarrow \boldsymbol{E(BY) = E(Y|B = 1)P(B = 1)}\big)$

$$= \int_{\overline{m}_k} \mathrm{E}(\bar{S}_k(a^*, \overline{m}_k)\, Y_{k+1}(a^*, 1, \overline{m}_k) | \overline{M}_k(a^*, \bar{s}_{k-1} = 1) = \overline{m}_k, L_0 = l_0) f(\overline{M}_k(a^*, \bar{s}_{k-1} = 1)$$

$$= \overline{m}_k | \bar{S}_{k-1} = 1, L_0 = l_0)\, d\overline{m}_k$$

$$= \int_{\overline{m}_k} \mathrm{E}(\bar{S}_k(a^*, \overline{m}_k)\, Y_{k+1}(a^*, 1, \overline{m}_k) | \overline{M}_k(a^*, \bar{s}_{k-1} = 1) = \overline{m}_k, A = a^*, L_0 = l_0) f(\overline{M}_k(a^*, \bar{s}_{k-1}$$

$$= 1) = \overline{m}_k | \bar{S}_{k-1} = 1, L_0 = l_0)\, d\overline{m}_k$$

$\because \boldsymbol{Y_{k+1}(a^*, \bar{s}_k = 1, \overline{m}_k) \perp A \,|L_0}$ **and** $\boldsymbol{\bar{S}_k(a^*, \overline{m}_k) \perp A \,|L_0}$

$$= \int_{\overline{m}_k} \mathrm{E}(\bar{S}_k(a^*, \overline{m}_k)\, Y_{k+1}(a^*, 1, \overline{m}_k) | A = a^*, L_0 = l_0) f(\overline{M}_k(a^*, 1)$$

$$= \overline{m}_k | \bar{S}_{k-1} = 1, L_0 = l_0)\, d\overline{m}_k$$

$\because \boldsymbol{\left(Y_{k+1}(a^*, \bar{s}_k = 1, \overline{m}_k), \bar{S}_k(a^*, \overline{m}_k)\right) \perp \overline{M}_k \,|L_0}$

$$= \int_{\overline{m}_k} \mathrm{E}(\bar{S}_k(a^*, \overline{m}_k)\, Y_{k+1}(a^*, 1, \overline{m}_k) | A = a^*, \overline{M}_k = \overline{m}_k, L_0 = l_0) f(\overline{M}_k(a^*, 1)$$

$$= \overline{m}_k | \bar{S}_{k-1} = 1, L_0 = l_0) d\overline{m}_k$$

$\because \boldsymbol{Y_{k+1}(a^*, \bar{s}_k = 1, \overline{m}_k) \perp \overline{M}_k | A, \bar{S}_{k-1} = 1, L_0}$ **and** $\boldsymbol{\bar{S}_k(a^*, \overline{m}_k) \perp \overline{M}_k | A, L_0}$

$$= \int_{\overline{m}_k} \mathrm{E}(\bar{S}_k Y_{k+1}(a^*, 1, \overline{m}_k) | A = a^*, \overline{M}_k = \overline{m}_k, L_0 = l_0) f(\overline{M}_k(a^*, 1)$$

$$= \overline{m}_k | \bar{S}_{k-1} = 1, L_0 = l_0) d\overline{m}_k$$

$\because \boldsymbol{consistency}$

$$= \int_{\overline{m}_k} \mathrm{E}(Y_{k+1}(a^*, 1, \overline{m}_k) | \bar{S}_k = 1, A = a^*, \overline{M}_k = \overline{m}_k, L_0 = l_0) f(\bar{S}_k$$

$$= 1 | A = a^*, \overline{M}_k = \overline{m}_k, L_0 = l_0) f(\overline{M}_k(a^*, 1) = \overline{m}_k | \bar{S}_{k-1} = 1, L_0 = l_0) d\overline{m}_k$$

$$= \int_{\overline{m}_k} \mathrm{E}(Y_{k+1}(a^*, 1, \overline{m}_k) | \bar{S}_k = 1, A = a^*, \overline{M}_k = \overline{m}_k, L_0 = l_0) f(\bar{S}_k$$

$$= 1 | A = a^*, \overline{M}_k = \overline{m}_k, L_0 = l_0) f(\overline{M}_k(a^*, 1)$$

$$= \overline{m}_k | A = a^*, \bar{S}_{k-1} = 1, L_0 = l_0) d\overline{m}_k$$

$\because \boldsymbol{\overline{M}_k(a^*, \bar{s}_{k-1} = 1) \perp A \,| \bar{S}_{k-1} = 1, L_0}$

$$= \int_{\overline{m}_k} \mathrm{E}(Y_{k+1} | A = a^*, \bar{S}_k = 1, \overline{M}_k = \overline{m}_k, L_0 = l_0) f(\bar{S}_k = 1 | A = a^*, \overline{M}_k = \overline{m}_k, L_0 = l_0) f(\overline{M}_k$$

$$= \overline{m}_k | A = a^*, \bar{S}_{k-1} = 1, L_0 = l_0) d\overline{m}_k$$

$\because$ \mathbit{**_consistency_**}

Please note that the proof for the direct effect and the indirect effect through death can be obtained analogously.